\documentclass[aps,prd,twocolumn,superscriptaddress]{revtex4-2}

\usepackage[dvips]{graphicx}
\usepackage{ulem}
\usepackage{float}
\usepackage[usenames]{color}
\usepackage{amsmath}
\usepackage{amssymb}
\usepackage{bm}
\usepackage{mathrsfs}
\usepackage{amsbsy}
\usepackage{hyperref}
\usepackage[all]{hypcap}
\usepackage{longtable}
\usepackage{url}
\usepackage{multirow}
\usepackage{booktabs}
\usepackage[utf8]{inputenc}
\usepackage{xcolor} 

\begin{document}

\title{Spectrum of \([cq][\bar{s}\bar{q}]\) tetraquarks: Nature of $D^*_{s0}(2317)$, $D_{s1}(2460)$ and $T^*_{c\bar s0}(2900)$}

\author{Zhi-Yuan Chen}
\author{Zhe-Hao Cao}
\author{You-You Lin}
\author{Ji-Ying Wang}
\author{Ailin Zhang}
\email{zhangal@shu.edu.cn}
\affiliation{Department of Physics, Shanghai University, Shanghai 200444, China}

\begin{abstract}
Motivated by the recent observations of exotic open-charm tetraquark candidates \(T^a_{c\bar{s}0}(2900)^{++}\) and \(T^a_{c\bar{s}0}(2900)^{0}\), we systematically calculate the mass spectra of \([cq][\bar{s}\bar{q}]\) tetraquarks within a nonrelativistic constituent quark potential model. In the model, the tetraquark states are treated as diquark-antidiquark bound systems with an interior interaction similar to the quark-antiquark interaction in conventional mesons. The well established states \(D_{s0}^*(2317)\) with \(J^P=0^+\) and \(D_{s1}(2460)\) with \(J^P=1^+\) could be identified as the two ground states of the \([cq][\bar{s}\bar{q}]\) system. \(T^a_{c\bar{s}0}(2900)^{0}\) and \(T^a_{c\bar{s}0}(2900)^{++}\) could be naturally interpreted as radially excited \(0^+\) tetraquark states with different interior components. Their large mass difference may result from their different interior structure instead of an isospin symmetry breaking. Whether \(T^a_{c\bar{s}0}(2900)^{0}\) and \(T^a_{c\bar{s}0}(2900)^{++}\) belong to an isospin triplet deserves further experimental investigation. In addition, there may be another \(0^+\) \([cq][\bar{s}\bar{q}]\) tetraquark state with mass around $2450$ MeV, which is composed of a $cq$ diquark and a $\bar s\bar q$ antidiquark both with spin-0. In the energy region $2640-2700$ MeV, there may be a $J^P=2^+$ \([cq][\bar{s}\bar{q}]\) tetraquark state composed of the $cq$ diquark and the $\bar s\bar q$ antidiquark both with spin-1.
\end{abstract}

\maketitle
\section{Introduction}\label{intro}

\(D^*_{s0}(2317)^\pm\) was first observed by the BaBar Collaboration~\cite{PhysRevLett.90.242001} and confirmed by the CLEO and BELLE Collaborations~\cite{PhysRevD.68.032002,prl92.012002}, but the neutral and doubly charged partners of \(D^*_{s0}(2317)^+\) have not been found yet. \(D_{s1}(2460)^\pm\) was first reported by the CLEO Collaboration~\cite{PhysRevD.68.032002} and also observed by the Belle and BaBar Collaborations~\cite{prl91.262002,prd69.031101}. The masses and widths of \(D^*_{s0}(2317)\) and \(D_{s1}(2460)\) from Particle Data Group's (PDG)~\cite{pdg24} average are
\begin{align*}
D^*_{s0}(2317)^\pm: M = 2318.0 \pm 0.7 \,\text{MeV},\\
                  \quad \Gamma < 3.8 \,\text{MeV},\notag \\
D_{s1}(2460)^\pm: M = 2459.6 \pm 0.9 \,\text{MeV},\\
                  \quad \Gamma < 3.5 \,\text{MeV}.\notag
\end{align*}

In addition to the interpretation of conventional charmed strange $D_s$ mesons, these two states were interpreted as four-quark states, $DK/\pi$ molecules or baryoniums, for their masses being significantly lower than theoretical predictions~\cite{prd68.054006,plb567.23,prd70.096011,prl93.232001,prl94.162002}.

Recently, two other open heavy flavor candidates, \(T^a_{c\bar{s}0}(2900)^{++}\) and \(T^a_{c\bar{s}0}(2900)^{0}\), were reported by the LHCb collaboration in the \(D_s^+\pi^\pm\) invariant mass spectra in the \(B^+ \to D^-D_s^+\pi^+\) and \(B^0 \to \overline{D}^0 D_s^+\pi^-\) decays\cite{aaij2023first,aaij2023amplitude}. They have masses and decay widths
\begin{align*}
T^a_{c\bar{s}0}(2900)^{++}: M = 2921 \pm 17 \pm 20 \,\text{MeV},\\
                  \quad \Gamma = 137 \pm 32 \pm 17 \,\text{MeV},\notag \\
T^a_{c\bar{s}0}(2900)^{0}: M = 2892 \pm 14 \pm 15 \,\text{MeV},\\
                  \quad \Gamma = 119 \pm 26 \pm 13 \,\text{MeV}.\notag
\end{align*}
Both states are assumed to have $J^P=0^+$ and suggested to be an isospin triplet. Their decays imply \(T^a_{c\bar{s}0}(2900)^{++}\) and \(T^a_{c\bar{s}0}(2900)^{0}\) with minimal quark contents $[c\bar su\bar d]$ and $[c\bar s\bar ud]$, respectively. It was believed the first observation of an isospin triplet with four different quark flavors.

Four-quark states have been studied for many years. Since the observation of $T^*_{c\bar s0}(2900)$, this state was interpreted as a tetraquark $T_{c\bar s}$ in a constituent-quark-model-based coupled-channels calculation~\cite{prd108.094035}. It was interpreted with a hadronic molecule $D^*K^*$~\cite{prd107.094019,prd109.034027}, a virtual state~\cite{prd108.074006}, or a threshold effect from the interaction of the $D^*K^*$ channel~\cite{epjc82.955,prd107.056015}. Most of these studies focus on the specific properties of one or two independent states.

In a nonrelativistic potential quark model, the mass spectrum of $1S$-wave charmed-strange tetraquark states $cn\bar s\bar n$ and $cs\bar n\bar n$ ($n=u$ or d) was calculated~\cite{prd107.096020}, where the lowest $1S$-wave $J^P=0^+$ $cn\bar s\bar n$ tetraquark state has mass $2828$ MeV.
In a quasi-potential Bethe-Salpeter equation approach together with the one-boson exchange model~\cite{prd104.094012}, some heavy-strange meson systems were systematically investigated, and $D^*_{s0}(2317)$, $D_{s1}(2460)$ and $X_0(2900)$ were related to the $DK$ ($J^P=0^+$), $D^*K$ ($J^P=1^+$) and $\bar D^*K^*$ ($J^P=0^+$) molecular states, respectively.

So far, the nature of $D^*_{s0}(2317)$, $D_{s1}(2460)$ and $T^*_{c\bar s0}(2900)$ has not been identified. In order to know more about these states, we calculate the masses of \([cq]\) ($q=u$ or $d$) diquark and \([\bar{s}\bar{q}]\) antidiquark first, and then systematically calculate the mass spectrum of the \([cq][\bar{s}\bar{q}]\) tetraquark states within a non-relativistic constituent quark potential model. Based on the numerical results, we give the explanations of these states.

In the practical calculation, a \([cq][\bar{s}\bar{q}]\) tetraquark state is assumed to be composed of a $cq$ diquark and a $\bar s\bar q$ antidiquark. The interaction between the diquark and the antidiquark is assumed as that between a quark and another quark except for a difference of the color factor. In this work, universal nonrelativistic Semay-Silvestre-Brac potentials~\cite{zpc61.271,fbs20.1} will be employed.

The article is structured as follows: A brief introduction is given in Sect. I and the construction of the potential model and the wave functions are introduced in Sect. II. Numerical calculation and the mass spectra are presented in Sect. III. A brief summary is given in the last section.

\section{Constituent diquark and antidiquark potential model}\label{model}

To calculate the mass spectrum of mesons and baryons, nonrelativistic potentials for the quark-antiquark interaction were proposed by Semay and Silvestre-Brac~\cite{zpc61.271,fbs20.1} in the following explicit form
\begin{align}\label{vqq}
V_{q\bar{q}} = -\frac{\alpha(1 - e^{-r/r_c})}{r} + \lambda r^p + C \nonumber  \notag \\
+ (\vec{s}_1 \cdot \vec{s}_2) \frac{8\kappa(1 - e^{-r/r_c}) e^{-r^2/r_0^2}}{3m_1 m_2 \sqrt{\pi} r_0^3},  \end{align}
where \(r_0\) is determined by
\begin{align}
r_0 = A \left( \frac{2m_1 m_2}{m_1 + m_2} \right)^{-B}.
\end{align}
The parameters \( r_c \) and \( p \) represent the variations of the potential compared to the Cornell potential at both short and long distances. Since the potential is proportional to the product of two Gell-Mann matrices \( \lambda_i \cdot \lambda_j \), the quark-quark potential is half of the quark-antiquark potential once the diquark is constrained in color antitriplet.
Therefore, the quark-quark interaction potential in baryons or multiquark states is obtained with \(V_{qq}=\frac{1}{2}V_{q\bar{q}}\).

Recently, this quark-antiquark interaction potential was also employed to calculate the mass spectrum of doubly charmed tetraquark states \(T_{cc}\)~\cite{PhysRevD.111.014015} and hidden charmed tetraquark states~\cite{prd112.034036}, where the interaction between the diquark and the antidiquark is assumed to have the same form as that between the quark and the antiquark. In this model, the diquarks and antidiquarks inside tetraquarks are treated as compact entities without radial or orbital excitation between the two quarks or the two antiquarks.

A universal Breit-Fermi approximation~\cite{PhysRevD.12.147}
\begin{align}\label{vsl}
V_{sl} = & \frac{\alpha(1-e^{-r / r_c})}{r^3}\left(\frac{1}{m_1}+\frac{1}{m_2}\right)\left(\frac{\vec{s}_1}{m_1}+\frac{\vec{s}_2}{m_2}\right) \cdot \vec{L} \nonumber \\
& -\frac{1}{2 r} \frac{\partial V_{\text {cornell}}}{\partial r}\left(\frac{\vec{s}_1}{m_1^2}+\frac{\vec{s}_2}{m_2^2}\right) \cdot \vec{L} \nonumber \\
& +\frac{1}{3 m_1 m_2}\left(\frac{1}{r} \frac{\partial V_{Coul}}{\partial r}-\frac{\partial^2 V_{Coul}}{\partial r^2}\right)\\
&\times\left(\frac{3 (\vec{s}_1 \cdot \vec{r}) (\vec{s}_2 \cdot \vec{r})}{r^2}-\vec{s}_1 \cdot \vec{s}_2\right)\nonumber
\end{align}
is employed for the spin-orbit coupling in tetraquarks, where \( V_{Cornell} = -\frac{\alpha(1 - e^{-r/r_c})}{r} + \lambda r^p \) and \( V_{Coul} \) represents the modified Coulomb potential. The explicit form of \( V_{sl} \) can be expressed as follows
\begin{align}\label{vsl2}
V_{sl} = & V_1 \vec{L} \cdot \vec{s}_1 + V_2 \vec{L} \cdot \vec{s}_2 \nonumber\\
& + V_t\left(\frac{3 (\vec{s}_1 \cdot \vec{r}) (\vec{s}_2 \cdot \vec{r})}{r^2}-\vec{s}_1 \cdot \vec{s}_2\right)\
\end{align}
with
\begin{align}
V_1 = & \frac{\alpha\left(1-e^{-r / r_c}\right)}{m_1 m_2 r^3} +\frac{1}{2 m_1^2}\left(\frac{\alpha\left(1-e^{-r / r_c}\right)}{r^3}-\frac{\alpha e^{-r / r_c}}{r^2 r_c}-\frac{p\lambda r^{p-1}}{r}\right),\nonumber\\
V_2 = & \frac{\alpha\left(1-e^{-r / r_c}\right)}{m_1 m_2 r^3} +\frac{1}{2 m_2^2}\left(\frac{\alpha\left(1-e^{-r / r_c}\right)}{r^3}-\frac{\alpha e^{-r / r_c}}{r^2 r_c}-\frac{p\lambda r^{p-1}}{r}\right),\nonumber\\
V_t= & \frac{\alpha}{m_1 m_2}\left(\frac{1-e^{-r / r_c}}{r^3}-\frac{e^{-r / r_c}}{r^2 r_c}-\frac{e^{-r / r_c}}{3 r r_c^2}    \right),\nonumber
\end{align}
where \( \vec{s}_1 \) and \( \vec{s}_2 \) denote the spin angular momentum of the diquark and the antidiquark, and \( \vec{L}\) stands for the orbital angular momentum between them.

The standard spin-orbit interaction derived from the one-gluon exchange potential contains a $1/r^3$ singularity at short distance, and this strong singularity leads to an unphysical divergence in numerical calculations. To regularize this behavior, a phenomenological form factor, $(1 - e^{-r/r_0})$, is introduced into the spin-dependent terms. Physically, this modification accounts for the finite-size effects of the constituent quarks, effectively smoothing out the short-range singularity while preserving the right long-range behavior of the potential.

The full Hamiltonian of a charmed tetraquark state is constructed as~\cite{PhysRevD.111.014015}

\begin{equation}
H = m_1 + m_2 + \frac{p^2}{2\mu} + V_{q\bar{q}} + V_{sl},
\end{equation}
where $m_1$ and $m_2$ are masses of the charmed diquark and the light antidiquark, respectively. $\frac{p^2}{2\mu}$ is the kinetic energy of the tetraquark in the center-of-mass system, determined by the relative momentum $p$ and the reduced mass $\mu = \frac{m_1 m_2}{m_1 + m_2}$. $V_{q\bar{q}}$ represents the Semay-Silvestre-Brac potentials as described in Eq.~(\ref{vqq}), and $V_{sl}$ denotes the spin-dependent potential given in Eq.~(\ref{vsl}).

There are $AL$-type and $AP$-type phenomenological potentials in Refs.~\cite{zpc61.271,fbs20.1}. The key difference among these potentials lies in two parameters: the power index $p$ in the confining term and the cutoff radius $r_c$ that regularizes the Coulomb-like singularity at short distances. That is to say

\begin{itemize}
    \item $AL1$ potential: $p = 1$, $r_c = 0$,
    \item $AL2$ potential: $p = 1$, $r_c \neq 0$,
    \item $AP1$ potential: $p = \frac{2}{3}$, $r_c = 0$,
    \item $AP2$ potential: $p = \frac{2}{3}$, $r_c \neq 0$.
\end{itemize}
The $AL$ potentials adopt a linear confining term ($p=1$), while the $AP$ potentials employ a power form ($p=2/3$). Since there is no excitation inside diquarks and antidiquarks, the $AL$ potentials are employed only for the calculation of their masses as that in Ref.~\cite{PhysRevD.111.014015}.

As a two body system, the wave function of the tetraquark state is obtained by solving the non-relativistic Schrödinger equation. The wave function of the $[cq][\bar{s}\bar{q}]$ tetraquark system is expressed as a direct product of flavor, color, spin, and spatial components:
\begin{align}
\Psi_{JM} = \chi_{\text{flavor}} \otimes \chi_{\text{color}} \otimes \bigl[\Psi_{lm}(\vec{r})\otimes\chi_{\text{spin}}\,\bigr]_{JM},
\end{align}
where $\Psi_{lm}(\vec{r})$ represents the spatial wave function with orbital angular momentum $l$, $\chi_{\text{spin}}$ denotes the spin wave function. The orbital angular momentum is coupled with the spin momentum to form the total angular momentum $J$ and its projection $M$. The diquark and the antidiquark in tetraquark states may be in a $\bar{3}_c \otimes 3_c$ color-configuration or a $\overline{6}_{c} \otimes 6_{c}$.

For the $[cq][\bar{s}\bar{q}]$ system, the flavor structure must be constructed with definite isospin quantum numbers. The light antiquarks can form either symmetric or antisymmetric flavor combinations, which was defined as $\{\bar{q}_3\bar{q}_4\} = \frac{1}{\sqrt{2}}(\bar{q}_3\bar{q}_4 + \bar{q}_4\bar{q}_3)$ and $[\bar{q}_3\bar{q}_4] = \frac{1}{\sqrt{2}}(\bar{q}_3\bar{q}_4 - \bar{q}_4\bar{q}_3)$. Obviously, the curly braces denote symmetric combinations and the square brackets denote antisymmetric combinations.

Following the isospin coupling rules in the $\bar{3}_F$ flavor representation, the $[cq][\bar{s}\bar{q}]$ system can form both isospin singlet ($I=0$) and triplet ($I=1$) states. For the isospin singlet state with $I=0$, $I_3=0$, the flavor wave function is given by $\frac{1}{\sqrt{2}}cd[\bar{s}\bar{d}] + \frac{1}{\sqrt{2}}cu[\bar{s}\bar{u}]$. For the isospin triplet states with $I=1$, there are three components corresponding to different $I_3$ values: flavor wave function $cu[\bar{s}\bar{d}]$ with $I_3=+1$, $\frac{1}{\sqrt{2}}cd[\bar{s}\bar{d}] - \frac{1}{\sqrt{2}}cu[\bar{s}\bar{u}]$ with $I_3=0$, and $cd[\bar{s}\bar{u}]$ with $I_3=-1$. Anyway, the isospin quantum numbers do not affect the final mass spectrum of tetraquark states in the diquark-antidiquark picture.

For the $\bar{3}_c \otimes 3_c$ color-configuration in $[cq][\bar{s}\bar{q}]$ tetraquark states, we have six independent color-spin configurations:
\begin{align}
\chi_{1} &= |(cq)_0^{\bar{3}} [\bar{s}\bar{q}]_0^{3} \rangle_0, \label{eq:colorspin1} \\[8pt]  
\chi_{2} &= |(cq)_0^{\bar{3}} \{\bar{s}\bar{q}\}_1^{3} \rangle_1, \label{eq:colorspin2} \\[8pt] 
\chi_{3} &= |(cq)_1^{\bar{3}} [\bar{s}\bar{q}]_0^{3} \rangle_1, \label{eq:colorspin3} \\[8pt]
\chi_{4} &= |(cq)_1^{\bar{3}} \{\bar{s}\bar{q}\}_1^{3} \rangle_0, \label{eq:colorspin4} \\[8pt]
\chi_{5} &= |(cq)_1^{\bar{3}} \{\bar{s}\bar{q}\}_1^{3} \rangle_1, \label{eq:colorspin5} \\[8pt]
\chi_{6} &= |(cq)_1^{\bar{3}} \{\bar{s}\bar{q}\}_1^{3} \rangle_2. \label{eq:colorspin6}
\end{align}
where the notation $|(cq)_{s_1}^{\bar{3}} (\bar{s}\bar{q})_{s_2}^{3} \rangle_S$ represents a diquark with spin $s_1$ in color-antitriplet $\bar{3}_c$, an antidiquark with spin $s_2$ in color-triplet $3_c$, and total spin $S$. In the above configurations, the square brackets and curly braces for the antidiquark represent antisymmetric and symmetric flavor wave functions, respectively. Explicitly, for the $I=0$ configurations, $cq\{\bar{s}\bar{q}\} \equiv (cd\{\bar{s}\bar{d}\} + cu\{\bar{s}\bar{u}\})/\sqrt{2}$ and $cq[\bar{s}\bar{q}] \equiv (cd[\bar{s}\bar{d}] + cu[\bar{s}\bar{u}])/\sqrt{2}$. For the $I=1$ configurations, $cq\{\bar{s}\bar{q}\} \equiv (cu\{\bar{s}\bar{d}\}, (cd\{\bar{s}\bar{d}\} - cu\{\bar{s}\bar{u}\})/\sqrt{2}, cd\{\bar{s}\bar{u}\})$ and $cq[\bar{s}\bar{q}] \equiv (cu[\bar{s}\bar{d}], (cd[\bar{s}\bar{d}] - cu[\bar{s}\bar{u}])/\sqrt{2}, cd[\bar{s}\bar{u}])$.

For tetraquarks in the $\mathbf{\overline{6}_{c} \otimes 6_{c}}$ color configuration, six kinds of color spin wave functions are theoretically allowed. However, the color interactions within such configurations exhibit repulsive characteristics \cite{JAFFE20051}. This repulsive nature suppresses the formation of compact correlated structures composed of diquarks and antidiquarks. So the configurations in the \(\mathbf{6_c}\) color representation will not be considered and the more physically favored $\mathbf{\bar{3}_c \otimes 3_c}$ configuration is focused on.

The Gaussian Expansion Method (GEM) \cite{HIYAMA2003223} is employed for the computation of the spatial wave function $\psi_{lm}(\vec{r})$, where the wave functions are expanded as a linear combination of Gaussian basis functions that serve as trial solutions in the Schrödinger formalism
\begin{align}
\Psi_{lm}(\vec{r}) &= \sum_{n=1}^{n_{max}} c_{nl} \phi_{nlm}(\vec{r}) \\[4pt]
\phi_{nlm}(\vec{r}) &= N_{nl} r^l e^{-r^2/r_n^2} Y_{lm}(\hat{r}).
\end{align}
Here, $N_{nl}$ denotes the normalization factor for states characterized by the radial quantum number $n$ and orbital angular momentum quantum number $l$. The Gaussian width parameters $r_n$ are obtained according to a geometric progression
\begin{align}
r_n = r_1 a^{n-1} \quad (n=1,2,...,n_{max}),
\end{align}
with three parameters $\{n_{max}, r_1, r_{n_{max}}\}$~\cite{HIYAMA2003223}.

By implementing the Rayleigh-Ritz variational principle, we convert the Schrödinger equation into a generalized eigenvalue problem expressed in matrix form
\begin{align}
\sum_{n=1}^{n_{max}} \sum_{n'=1}^{n_{max}} (H_{nn'} - E_{nl}N_{nn'})c_{nl} = 0.
\end{align}
The Hamiltonian matrix elements $H_{nn'}$ and overlap integrals $N_{nn'}$ are defined as
\begin{align}
H_{nn'} &= \langle\phi_{n'lm}(\vec{r})|H|\phi_{nlm}(\vec{r})\rangle, \\[8pt]
N_{nn'} &= \langle\phi_{n'lm}(\vec{r})|\phi_{nlm}(\vec{r})\rangle.
\end{align}

Both the expansion coefficients $c_{nl}$ and the corresponding energy eigenvalues $E$ are then yielded. To achieve numerical convergence, we initialize our calculations with $r_0 = 0.1$ fm, which has negligible impact on the final results. Following methodologies established in previous studies~\cite{HIYAMA2003223,PhysRevD.105.054015,prd112.034036}, we optimize the parameters $r_n$ and $n_{max}$ accordingly and set the variational parameters $\{r_1, r_{n_{max}}, n_{max}\}$ to {0.1 fm, 5 fm, 25} for the geometric progression.

\section{Numerical results}\label{results}

In this paper, the constituent quark masses and quark potential parameters are adopted from Refs.~\cite{zpc61.271,fbs20.1} where both mesons and baryons mass spectra were successfully reproduced. The complete parameters are presented in Table I.

\begin{table}[H]  
\centering
\caption{Parameters in four \textit{AL} and \textit{AP} potentials.}
\label{tab:ALparameters}
\begin{tabular*}{0.5\textwidth}{@{\extracolsep{\fill}} l c c c c @{}}
\hline\hline
\textbf{Parameters} & \textbf{AL1} & \textbf{AL2} & \textbf{AP1} & \textbf{AP2} \\
\hline
$m_{u,d}(\mathrm{GeV})$   & 0.315  & 0.320  & 0.277  & 0.280  \\
$m_{c}(\mathrm{GeV})$     & 1.836  & 1.851  & 1.819  & 1.840  \\
$m_{s}(\mathrm{GeV})$     & 0.577  & 0.587  & 0.553  & 0.569  \\
$\alpha$                  & 0.5069 & 0.5871 & 0.4242 & 0.5743 \\
$\kappa$                  & 1.8609 & 1.8475 & 1.8025 & 1.8093 \\
$\lambda$                 & 0.1653 & 0.1673 & 0.3898 & 0.3972 \\
$C(\mathrm{GeV})$         & $-0.8321$ & $-0.8182$ & $-1.1313$ & $-1.1146$ \\
$B$                       & 0.2204 & 0.2132 & 0.3263 & 0.3478 \\
$A(\mathrm{GeV}^{B-1})$   & 1.6553 & 1.6560 & 1.5296 & 1.5321 \\
$r_{c}(\mathrm{GeV}^{-1})$& 0      & 0.1844 & 0      & 0.3466 \\
\hline\hline
\end{tabular*}
\end{table}

As a point-like objects without excitation, a diquark may have spin-$0$ or spin-$1$. Therefore, two AL-type potentials are employed to calculate the masses of diquark and antidiquark. The masses of $cq$ diquarks and $\bar s\bar q$ antidiquark with different spins are then calculated and presented in Table II.

The masses of \(cq\) diquark are the same as those in Ref.~\cite{PhysRevD.111.014015}. The scalar charm-light diquark with spin-$0$ has mass $\sim 310$ MeV larger than the predicted one in QCD sum rule~\cite{prd87.125018}, and the vector charm-light diquark with spin-$1$ has mass $\sim 340$ MeV larger than the predicted one in QCD sum rule. The mass of $\bar s\bar q$ antidiquark is almost twice as large as that ($\sim 460$ MeV) predicted in QCD sum rule~\cite{prd76.036004}.
The scalar diquark has a lower mass in comparison to the vector diquark, which is consistent with existed theoretical predictions and phenomenological analyses. The mass difference between the scalar and the vector $cq$ diquark is about $40$ MeV, and the mass difference between the scalar and the vector $\bar s\bar q$ diquark is about $110$ MeV. Obviously, the spin-related interaction has large effect on the masses of light diquarks and antidiquarks. The mass difference between two AL-type potentials is about $10-15$ MeV, so the diquark masses from $AL1$ potential will be used only in the following calculation.

\begin{table}[H]  
  \centering
  \caption{Masses (in MeV) of the diqurak and antidiquark.}
 \label{tab:ALparameters}
\begin{tabular*}{0.5\textwidth}{@{\extracolsep{\fill}} l c c  c @{}}
    \hline\hline
    \textbf{} & \textbf{spin}  & \textbf{AL1} & \textbf{AL2}   \\
    \hline

    \multirow{2}{*}{${cq}$} & $s = 0$  & 2169.63 & 2184.68   \\
                              & $s = 1$  & 2211.89 & 2227.49   \\

    \multirow{2}{*}{${\bar{s}\bar{q}}$} & $s = 0$  & 915.31 & 926.75   \\
                              & $s = 1$  & 1026.15 & 1037.72   \\

    \hline\hline
  \end{tabular*}
\end{table}

In terms of the parameters in Table I, the mass spectra of the ground \([cq][\bar{s}\bar{q}]\) tetraquarks are about $150$ MeV higher than the masses of \(D_{s0}^*(2317)\) and \(D_{s1}(2460)\). The distance between a diquark and an
antidiquark is likely larger than the distance between a quark and an antiquark in ordinary mesons, the interaction parameters $\alpha$ and $\lambda$ in the quark-antiquark potential between the charmed diquark and the light antidiquark are possibly differ from those in ordinary mesons. Similar to the prescription in Ref.~\cite{PhysRevD.111.014015}, the parameters \(\alpha\) and \(\lambda\) are refitted through the two well established states $0^+$ \(D_{s0}^*(2317)\) and $1^+$ \(D_{s1}(2460)\), which are assumed two $cq\bar{s}\bar{q}$ tetraquark states. Since there are four different quark potentials between the diquark and the antidiquark, four sets of fitted parameters \(\alpha\) and \(\lambda\) are obtained and presented in Table III. Other parameters \(\kappa, C, B, A,\) and \(r_c\) are the same as the original sets~\cite{zpc61.271,fbs20.1}. The fitted masses for these two states in the four potentials vary negligible, with fluctuations less than $0.2$ MeV, which demonstrates the stability of the fitting procedure.

\begin{table}[H]  
\centering
\caption{Parameters $\alpha$ and $\lambda$ in different potentials.}
\label{tab:ALparameters}
\begin{tabular*}{0.5\textwidth}{@{\extracolsep{\fill}} l c c c c @{}}
\hline\hline
\textbf{Potentials} & \textbf{AL1} & \textbf{AL2} & \textbf{AP1} & \textbf{AP2} \\
\hline
$\alpha$                  & 0.6865 & 0.8121 & 0.3921 & 0.5634 \\
$\lambda$                 & 0.1581 & 0.1857 & 0.2550 & 0.2985 \\
\hline\hline
\end{tabular*}
\end{table}

From the interior components in \(cq\bar{s}\bar{q}\) tetraquarks, the lowest state with spin$-1$ $cq$ diquark and $\bar s\bar q$ antidiquark has $J^P=0^+$, the lowest state with a spin$-0$ $cq$ diquark and $\bar s\bar q$ antidiquark also has $J^P=0^+$, and the lowest state with a spin$-0$ diquark/antidiquark and a spin$-1$ antidiquark/diquark has $J^P=1^+$. Once $0^+$ \(D_{s0}^*(2317)\) and $1^+$ \(D_{s1}(2460)\) are assumed the \( 1 \lvert [1,1]_0, 0 \rangle_0 \) and \( 1 \lvert [1,1]_1, 0 \rangle_1 \) \(cq\bar{s}\bar{q}\) tetraquarks, respectively, for the $J^P$ quantum numbers, the mass spectra of \(cq\bar{s}\bar{q}\) tetraquarks with different diquark/antidiquark from $1S$ to $2P$ excitations are computed and presented in Tables IV-VI. In these tables, the first column lists the permitted tetraquark states from $1S$ to $2P$ wave excitations, along with their spectroscopic notation \( n \lvert [s_d, s_{\bar d}]_S, L \rangle_J \)~\cite{prd98.094015,prd112.034036} in the LS coupling scheme. The second column provides the spin-parity quantum numbers \(J^P\), with parity \(P = (-1)^L\). The masses from four different types of potentials are presented in the third to the fifth columns.

\begin{table}[ht]
\centering
\caption{Mass spectra (in MeV) of \(cq\bar{s}\bar{q}\) tetraquark composed of the \(cq\) diquark and the \(\bar{s}\bar{q}\) antidiquark both with spin-1. }
\label{spin1}
    \begin{tabular*}{0.5\textwidth}{@{\extracolsep{\fill}} lcccccc @{}}
        \toprule
        \( n \lvert [s_d, s_{\bar d}]_S, L \rangle_J \) & \( J^P \) & AL1 & AL2 & AP1 & AP2 \\ \midrule
        \( 1 \lvert [1,1]_0, 0 \rangle_0 \) & \( 0^+ \) & 2317.87 & 2317.85 & 2317.80 & 2317.86\\
        \( 1 \lvert [1,1]_1, 0 \rangle_1 \) & \( 1^+ \) & 2459.58 & 2459.54 & 2459.50 & 2459.55\\
        \( 1 \lvert [1,1]_2, 0 \rangle_2 \) & \( 2^+ \) & 2687.52 & 2699.64 & 2646.08 & 2666.70\\

        \textcolor{blue}{\( 2 \lvert [1,1]_0, 0 \rangle_0 \)} & \textcolor{blue}{\( 0^+ \)} & 3114.45 & 3220.74 & \textcolor{blue}{2923.78} & 3016.17\\
        \( 2 \lvert [1,1]_1, 0 \rangle_1 \) & \( 1^+ \) & 3149.01 & 3255.92 & 2957.15 & 3048.58\\
        \( 2 \lvert [1,1]_2, 0 \rangle_2 \) & \( 2^+ \) & 3227.35 & 3334.08 & 3031.49 & 3123.38\\
        \bottomrule
    \end{tabular*}
\end{table}

\begin{table}[ht]
\centering
\caption{Mass spectra (in MeV) of \(cq\bar{s}\bar{q}\) tetraquark composed of the \(cq\) diquark and the \(\bar{s}\bar{q}\) antidiquark both with spin-0. }
\label{spin0}
    \begin{tabular*}{0.5\textwidth}{@{\extracolsep{\fill}} lcccccc @{}}
        \toprule
        \( n \lvert [s_d, s_{\bar d}]_S, L \rangle_J \) & \( J^P \) & AL1 & AL2 & AP1 & AP2 \\ \midrule
        \( 1 \lvert [0,0]_0, 0 \rangle_0 \) & \( 0^+ \) & 2463.79 & 2472.25 & 2437.24 & 2449.00 \\
        \( 1 \lvert [0,0]_0, 1 \rangle_1 \) & \( 1^- \) & 2913.42 & 2993.99 & 2746.86 & 2817.87 \\

        \textcolor{blue}{\( 2 \lvert [0,0]_0, 0 \rangle_0 \)} & \textcolor{blue}{\( 0^+ \)} & 3071.72 & 3183.87 & \textcolor{blue}{2866.97} & 2961.46 \\
        \( 2 \lvert [0,0]_0, 1 \rangle_1 \) & \( 1^- \) & 3359.66 & 3511.44 & 3057.61 & 3179.89 \\
        \bottomrule
    \end{tabular*}
\end{table}

\begin{table}[ht]
\centering
\caption{Mass spectra (in MeV) of \(cq\bar{s}\bar{q}\) tetraquark composed of the \(cq\) diquark and the \(\bar{s}\bar{q}\) antidiquark with one spin-0 and one spin-1.}
\label{spin10}
    \begin{tabular*}{0.5\textwidth}{@{\extracolsep{\fill}} lcccccc @{}}
        \toprule
        \( n \lvert [s_d, s_{\bar d}]_S, L \rangle_J \) & \( J^P \) & AL1 & AL2 & AP1 & AP2 \\ \midrule
        \( 1 \lvert [1,0]_1, 0 \rangle_1 \) & \( 1^+ \) & 2503.83 & 2511.98 & 2478.02 & 2489.52\\
        \( 1 \lvert [0,1]_1, 0 \rangle_1 \) & \( 1^+ \) & 2543.56 & 2547.89 & 2527.51 & 2535.70\\

        \( 1 \lvert [1,0]_1, 1 \rangle_0 \) & \( 0^- \) & 2918.93 & 2979.54 & 2777.86 & 2836.10\\
        \( 1 \lvert [0,1]_1, 1 \rangle_0 \) & \( 0^- \) & 2965.52 & 3004.83 & 2823.09 & 2887.90\\

        \( 1 \lvert [1,0]_1, 1 \rangle_1 \) & \( 1^- \) & 2937.49 & 3009.20 & 2782.93 & 2847.85\\
        \( 1 \lvert [0,1]_1, 1 \rangle_1 \) & \( 1^- \) & 2983.44 & 3044.25 & 2828.49 & 2898.31\\

        \( 1 \lvert [1,0]_1, 1 \rangle_2 \) & \( 2^- \) & 2967.68 & 3054.73 & 2791.91 & 2867.70\\
        \( 1 \lvert [0,1]_1, 1 \rangle_2 \) & \( 2^- \) & 3001.97 & 3084.01 & 2837.70 & 2905.99\\

        \( 2 \lvert [1,0]_1, 0 \rangle_1 \) & \( 1^+ \) & 3111.41 & 3223.20 & 2907.49 & 3001.73\\
        \( 2 \lvert [0,1]_1, 0 \rangle_1 \) & \( 1^+ \) & 3147.08 & 3254.38 & 2953.73 & 3044.70\\

        \( 2 \lvert [1,0]_1, 1 \rangle_0 \) & \( 0^- \) & 3371.74 & 3508.45 & 3090.38 & 3203.80\\
        \( 2 \lvert [0,1]_1, 1 \rangle_0 \) & \( 0^- \) & 3407.83 & 3523.89 & 3125.68 & 3245.82\\

        \( 2 \lvert [1,0]_1, 1 \rangle_1 \) & \( 1^- \) & 3386.39 & 3531.38 & 3094.35 & 3212.34\\
        \( 2 \lvert [0,1]_1, 1 \rangle_1 \) & \( 1^- \) & 3422.79 & 3555.14 & 3131.70 & 3255.80\\

        \( 2 \lvert [1,0]_1, 1 \rangle_2 \) & \( 2^- \) & 3410.46 & 3566.98 & 3101.38 & 3226.91\\
        \( 2 \lvert [0,1]_1, 1 \rangle_2 \) & \( 2^- \) & 3438.59 & 3589.63 & 3145.39 & 3263.54\\
        \bottomrule
    \end{tabular*}
\end{table}

From Tables~\ref{spin1}-\ref{spin10}, there are six S-wave \(cq\bar{s}\bar{q}\) tetraquarks without radial excitation, and there are fourteen P-wave \(cq\bar{s}\bar{q}\) tetraquarks without radial excitation. Among the S-wave tetraquarks, there are two $0^+$ tetraquark states with a mass difference $\sim 120$ MeV.
The masses of S-wave states fall within the range of $2318-2688$ MeV, which is also the mass region for conventional P-wave or first radial S-wave $D_s$ mesons. The overlap of normal mesons with tetraquarks allows more extra mesons in this mass range and will make the experimental identification of a state difficult.

Both the lowest \(cq\bar{s}\bar{q}\) tetraquark with two spin-1 and two spin-0 diquark/antidiquark have quantum numbers $J^P=0^+$, and the tetraquark with spin-1 components has a mass $\sim 120-150$ MeV lower. The splitting among the S-wave multiplet varies from $140$ MeV to $370$ MeV, which is much larger than the splitting among conventional meson multiplets.

$D_{sJ}(2632)^+$ was first observed by SELEX Collaboration~\cite{prl93.242001}, and was interpreted as the radial excitation $D_s(2 ^3S_1)^+$ of the vector ground-state $D^{*+}_s$ meson. This state has not been observed by BABAR, FOCUS or Belle~\cite{prd80.071502r}. $D_{s0}(2590)^+$ was first observed by LHCb Collaboration~\cite{prl126.122002}, and suggested as the radial excitation $D_s(2 ^1S_0)^+$ of the pseudoscalar ground-state $D^+_s$ meson. According to Table ~\ref{spin1}, there is a $J^P=2^+$ \(cq\bar{s}\bar{q}\) tetraquark with mass around $2640-2700$ MeV. The observation and confirmation of charmed strange meson in this mass range will discover more information in hadron structure and dynamics.

Once \(T^a_{c\bar{s}0}(2900)^{0}\) and \(T^a_{c\bar{s}0}(2900)^{++}\) are confirmed the \(0^+\) tetraquark states with $cd\bar s\bar u$ and $cu\bar s\bar d$ flavor components, they are possibly the radially excited \(0^+\) tetraquark states corresponding to the configurations \(2\lvert [0,0]_0, 0 \rangle_0\) and \(2 \lvert [1,1]_0, 0 \rangle_0\) according to Table~\ref{spin1} and Table-\ref{spin0}. Around $2900$ MeV, some P-wave $0^-$, $1^-$ or $2^-$, or radially excited S-wave $1^+$ tetraquark states may exist in different potentials, these states will enrich the zoo at this energy region.

For conventional mesons, typical isospin symmetry breaking may bring in several MeV mass difference (with maximum limit to 10 MeV) to an isospin multiplet. In Refs.~\cite{prl113.112001,plb749.454,prd110.094025}, an isospin-dependent interaction has been introduced to the hidden-charm tetraquark states. This interaction brings in less than $10$ MeV to the mass spectra. This isospin-dependent interaction provides a dynamical explanation of the isospin breaking or mixing effects.

Both \(T^a_{c\bar{s}0}(2900)^{0}\) and \(T^a_{c\bar{s}0}(2900)^{++}\) have spin-parity $J^P=0^+$, and their resonant parameters are consistent with each other. It is natural to suggest that they belong to an isospin triplet. Superficially, the large mass splitting ($\sim 30$ MeV) between \(T^a_{c\bar{s}0}(2900)^{0}\) and \(T^a_{c\bar{s}0}(2900)^{++}\) may result from the isospin symmetry breaking effect. In fact, different interior structure with the same $J^P$ quantum numbers and quark components in tetraquark states may bring in dozens of MeV difference to their masses. From Tables~\ref{spin1} and ~\ref{spin0}, the \(2\lvert [0,0]_0, 0 \rangle_0\) composed of the \(cq\) diquark and the \(\bar{s}\bar{q}\) antidiquark both with spin-0 has a mass $\sim 60$ MeV larger than the \(2 \lvert [1,1]_0, 0 \rangle_0\) composed of the \(cq\) diquark and the \(\bar{s}\bar{q}\) antidiquark both with spin-1. For real physical states with possible mixing effect or other uncertainties, the large mass splittings $\sim 30$ MeV between \(T^a_{c\bar{s}0}(2900)^{0}\) and \(T^a_{c\bar{s}0}(2900)^{++}\) is possible once they are assumed as two tetraquark states with different interior structure. In short, the large mass difference between \(T^a_{c\bar{s}0}(2900)^{0}\) and \(T^a_{c\bar{s}0}(2900)^{++}\) results likely from their interior components instead of the isospin symmetry breaking effect.

For each \( n \lvert [s_d, s_{\bar d}]_S, L \rangle_J \) tetraquark, four different potentials may bring several tens of MeV difference to the S-wave state, $100\to 250$ MeV difference to the P-wave state, and $100-300$ MeV difference to the radial excitation. Which potentials is preferred for the tetraquark states may depend on the color screening effect and the finite-size effect of the diquarks.

Intuitively, the screening effect (screening factor $(1 - e^{-r/r_c})$) is expected to lower the energy eigenvalues by softening the long-range behavior of the potential. However, the masses of the radial excitations in AP2 potential are approximately $80$ MeV higher than those in AP1 potential. The observed upward mass shift can be attributed to the non-trivial interplay between the screening mechanism and the parameters fitting constrained by the two well identified states. The two lowest  state masses are fixed to the experimental $D_{s0}^*(2317)$ and $D_{s1}(2460)$ across all potentials, the introduction of the screening factor requires a compensatory enhancement of the effective coupling strength (or confinement parameters) in the short-to-intermediate range to counteract the energy reduction caused by screening. This parameter refitting results in a potential that is effectively stiffer in the intermediate radial region, which is significantly probed by the radially excited states. As a consequence, this compensatory strengthening of the interaction dominates over the long-range screening suppression for the low-lying excited states, and leads to the observed mass increase.

In our opinion, the softer power-law potential ($p=2/3$) may offer a more physical description to the radially excited tetraquarks. This preference arises from two competitive physical effects, the color screening effect and the finite-size effect of the diquarks, which become prominent as the spatial extent of the system increases.

First, the strict linear confinement ($V \propto r$) is valid primarily in the intermediate distance region to the typical low-lying heavy quarkonium. However, for radially excited states, the creation of light quark-antiquark pairs from the vacuum would screen the color charge which leads to a softening or saturation of the confining potential~\cite{li2009higher,bai2009bottomonium}. At these larger distances, unlike the linear potential which grows indefinitely, the effective interaction in QCD is expected to flatten out, and the power-law potential with $p=2/3$ serves as an effective phenomenological description of the screening effect. the average separation between the diquark and antidiquark increases significantly. This mechanism may work also to the radially excited tetraquark states.

Second, the scalar and vector diquarks are in fact composite objects with a finite size and internal structure~\cite{ebert2006masses,selem2006hadron} though they are treated as point-like objects in this paper. When the diquark-antidiquark separation is comparable to the intrinsic size of the diquarks (estimated by the parameter $r_0$ ), the interaction cannot be treated simply as a flux tube between two points. The spatial extension of the diquarks introduces a smearing effect to the gluon exchange and confinement forces. This structural complexity implies that the effective potential between two composite color sources should be softer than the potential between two point sources.

Historically, the efficency of potentials with exponents $p<1$ has been well established in heavy hadron spectroscopy. Martin~\cite{martin1980fit} has demonstrated that such a power-law potential often yield superior descriptions of heavy quarkonia spectra in comparison to the rigid linear plus colour Coulomb potential. For  $cq\bar{s}\bar{q}$ tetraquarks, the adoption of $p=2/3$ for radially excited states aligns with these classical observations, and may effectively get the dynamics of a loosely bound system subject to both vacuum polarization and structural smearing.

It is well known that the mixing of tetraquarks with conventional mesons, coupled-channel effects or other effects may shift predicted masses of tetraquarks to observed ones, especially for states near thresholds like $D_{s0}^*(2317)$. It is useful to estimate the theoretical uncertainty from the input parameters $\alpha$ and $\lambda$. Following Ref.~\cite{PhysRevD.111.014015}, we performed an uncertainty analysis by shifting the input masses of the assumed tetraquarks, $D_{s0}^*(2317)$ and $D_{s1}(2460)$, upward by $50$ MeV. The refitted parameters show a moderate variation: $\alpha_s$ decreases from $0.3920$ to $0.3659$, while the confinement strength $\lambda$ increases from $0.2550$ to $0.2757$. This uncertainty will change the the masses of ground states and orbital excitations by approximately $50 \sim 60$ MeV. The radial excitations (candidates for $T_{c\bar{s}0}(2900)$) show a slightly larger shift of $\sim 70$ MeV (e.g., $2866 \to 2934$ MeV). This behavior is physically expected, as spatially extended radial excitations are more sensitive to the increased confinement parameter $\lambda$. Anyway, this global systematic uncertainty of $\sim 50-70$ MeV preserves the mass hierarchy and splitting patterns for tetraquark states.

\section{Summary}\label{summary}

In this work, we study the open-charm tetraquark states by treating them as a combined system of a diquark and an antidiquark. The interactions between a quark and another quark, and the interactions between a diquark and an antidiquark are described by the Semay and Silvestre-Brac potentials. With an explicit construction of the wave functions for open-charm $cq\bar{s}\bar{q}$ tetraquark states, the mass spectra of open-charm tetraquark states from $1S$-wave excitations to $2P$-wave excitations are systemically estimated after some diquark masses have been estimated in the same way.

\(D_{s0}^*(2317)\) and \(D_{s1}(2460)\) could be interpreted the $0^+$ \( 1 \lvert [1,1]_0, 0 \rangle_0 \) and $1^+$ \( 1 \lvert [1,1]_1, 0 \rangle_1 \) \(cq\bar{s}\bar{q}\) tetraquarks, respectively. Accordingly, the parameters $\alpha$ and $\lambda$ are fixed by these two well established states. In addition, a $0^+$ ground \(cq\bar{s}\bar{q}\) tetraquark state around $2450$ MeV may exist, while the next $0^+$ conventional $D_s$ ($2~^3P_0$) meson is predicted to have mass above $3000$ MeV.

The two $0^+$ exotic candidates with four different flavors, $T^a_{c\bar{s}0}(2900)^{0}$ and $T^a_{c\bar{s}0}(2900)^{++}$, could be interpreted as the radially excited $[cq][\bar{s}\bar{q}]$ tetraquark states with configurations $2| [0,0]_0, 0 \rangle_0$ and $2 | [1,1]_0, 0 \rangle_0$, respectively. It is highly probable that the mass difference between these two states results from their different internal structures rather than isospin symmetry breaking. To distinguish the mass difference induced by internal structure from that caused by isospin symmetry breaking, further experimental measurements of their decay channels and relevant branching ratios are crucial. A $2^+$ ground \(cq\bar{s}\bar{q}\) tetraquark state around $2640-2700$ MeV and other \(cq\bar{s}\bar{q}\) tetraquark states with different $J^P$ quantum numbers around $2900$ MeV may exist.

While the linear potential ($p=1$) adequately describes the compact ground tetraquark states, the softer power-law potential ($p=2/3$) may be essential for reproducing realistic masses for radially excited tetraquark states. This distinction reflects the physical reality of color screening and the finite size of composite diquarks in higher excited states.

Our analysis has not taken into account the explicit contributions from configuration mixing and many-body interactions in open-charmed tetraquarks. Furthermore, isospin-dependent interactions or boson-exchange interactions~\cite{epjp139.707} have not been included. It should be noted that the parameters in the quark potentials, determined from $D_{s0}^*(2317)$ and $D_{s1}(2460)$, could introduce some uncertainties into the predicted mass spectrum. However, a straightforward estimation suggests that the resulting uncertainties in the mass predictions remain relatively small. As additional open-charmed tetraquarks are discovered and unambiguously identified, the corresponding experimental data can be employed to further refine these predictions.

\begin{acknowledgments}
This work is supported by National Natural Science Foundation of China under the grant No. 11975146.
\end{acknowledgments}

%

\end{document}